\documentclass[twocolumn]{webofc}
\usepackage{amsmath, amsthm, amssymb}
\usepackage{enumerate}
\usepackage{graphicx}
\graphicspath{{images/}}
\usepackage{physics}

\usepackage[varg]{txfonts}   % Web of Conferences font
\begin{document}
\title{Modeling particle-fluid interaction in a coupled CFD-DEM framework}

\author{
	\firstname{Ilberto} \lastname{Fonceca}\inst{1}\fnsep\thanks{\email{jfonceca@unav.es}} \and
        \firstname{Diego} \lastname{Maza}\inst{1} \and
        \firstname{Ra\'ul} \lastname{Cruz Hidalgo}\inst{1} 
}

\institute{Department of Physics and Applied Mathematics, Universidad de Navarra, Pamplona, Spain
          }

\abstract{
  In this work, we present an alternative methodology to solve the particle-fluid interaction in the 
  \textit{resolved} CFDEM{\textregistered}coupling framework. 
  This numerical approach consists of coupling a Discrete Element Method (DEM) with a 
  Computational Fluid Dynamics (CFD) scheme, solving the motion of immersed particles in a fluid phase. 
  As a novelty, our approach explicitly accounts for the body force acting on the fluid phase 
  when computing the local momentum balance equations.   
  Accordingly, we implement a fluid-particle interaction computing the buoyant and drag forces
  as a function of local shear strain and pressure gradient.
  As a benchmark, we study the Stokesian limit of a single particle.
  The validation is performed comparing our outcomes with the ones provided by a previous \textit{resolved}
  methodology and the analytical prediction. 
  In general, we find that the new implementation reproduces with very good accuracy the Stokesian dynamics. Complementarily, we study the settling terminal velocity of a sphere under confined conditions.
}
\maketitle
\section{Introduction}\label{sec-intro}

The recent advances in particle-fluid simulations allowed the description of many industrial and natural processes 
\cite{goniva_2012, hager_2012, feng_2003,ellero_2020}. For instance,  
the sedimentation of particles in dilute and dense conditions and the flow of particles through submerged 
hoppers  have been recently examined \cite{goniva_2012, hager_2012, durian_2017}.  
In general, experimental and simulated data agreed very 
well when exploring intermediate Reynolds number regimes.
It suggests that the CFD-DEM coupling can resolve the hydrodynamics interactions in those scales with enough accuracy. 

In the field of particle suspension modeling, researchers have been able to reproduce classical results,
applying a variety of methods combining Eulerian and Lagrangian approaches \cite{goniva_2012, hager_2012, zhao_2014}. 
When examining simulation domains that are much larger than the dimensions of the particle, \textit{unresolved 
methods} are typically applied.
In these methods, the local interactions are determined by auxiliary fields, which represent the particles in the 
CFD domain.  In contrast, in \textit{resolved methods}, the individual particles are fully discretized, and their 
dynamics are addressed with better accuracy. They are often used for smaller system sizes. 

The \textit{resolved} coupled CFD-DEM approach described in this paper is an extension of a previous Fictitious Domain Method (cfdemSolverIB) \cite{hager_2012}. In this implementation, there is no body force acting on the fluid, 
which implies that the local pressure profile is solely determined by the local variations of the velocity field. 
Accordingly, to account for the buoyancy acting on the particle due to the presence of the fluid, 
an exact term ($\rho_f g V_i$) is added, explicitly. 

In our approach, the gravitational field is directly included in the incompressible Navier-Stokes equation 
as a body force term. Consequently, the pressure profile also contains the gravitational contribution. That is why we no longer 
need to impose an explicit buoyant force acting on the particles. 
Thus, our approach does rely on the assumption that the use of absolute pressure to calculate the particle dynamical response 
is relevant when numerical corrections must be implemented at this scale. Although not relevant differences will be expected 
when a single particle is analyzed, such differences might be relevant for very dense suspensions where the local pressure 
varies due to the dynamical coupling between particles.

%Thus, our approach does not rely on the assumption that the impact of the body force (acting on the solid phase and liquid phase) is uniform  throughout the fluid phase. This might be relevant for very dense suspensions, where the local pressure varies with depth due to the presence of overburden  interacting particles.}

The present work is structured as follows: in Sections~\ref{subsec-fluid-phase} 
and~\ref{subsec-DEM}, we briefly explain the used CFD-DEM scheme. 
Moreover, in Section~\ref{subsec-particle_near_wall}, 
we briefly comment on a theoretical framework that describes the settling of a sphere in confined conditions. 
Finally, in Section~\ref{sec-results}, we present some numerical results highlighting the validity of our methodology.

\section{Model Description}\label{sec-model-desc}

\subsection{Fluid phase}\label{subsec-fluid-phase}

The governing equation for the fluid phase is the incompressible Navier-Stokes equation:
\begin{equation}\label{eq:inc-Navier}
 \rho_f \frac{d \vb{U}}{dt} + \rho_f \left( \vb{U} \vdot \grad \right) \vb{U} = 
 -\grad{p} + \mu \laplacian{\vb{U}} + \rho_f \vb{g} \, .
\end{equation}
Our approach explicitly considers the body force term, which was not considered in the previous implementations 
\cite{goniva_2012, hager_2012, hager_2014}.

When discretized for a CFD time step $\Delta t$, equation~\ref{eq:inc-Navier} becomes:

\begin{equation}\label{eq:discretized-Navier}
\begin{split}
 \rho_f & \frac{\vb{\hat{U}} - \vb{U}^{t - \Delta t}}{\Delta t} + 
 \rho_f \left(\vb{U}^{t - \Delta t} \vdot \grad \right) \vb{U}^{t - \Delta t} = \\ \\
 & - \grad{\hat{p}} + 
 \mu \laplacian \vb{U}^{t - \Delta t} +
 \rho_f \vb{g}
\end{split} 
\end{equation}
where $\hat{p}$ is the estimated pressure, $\vb{\hat{U}}$ the interim solution for $\vb{U}$ and 
$\vb{U}^{t - \Delta t}$ is the solution found in the previous time step. Likewise, the continuity equation holds:

\begin{equation}\label{eq:continuity}
 \div{\vb{U}} = 0 \, .
\end{equation}

For each particle, we determine its occupation region $\mathcal{T}_i$ in the CFD grid by applying a correction to the velocity 
field in all the corresponding cells. In the cell $c \in \mathcal{T}_i$, we retrieve the velocity $\vb{\hat{U}}_c$ and 
pressure $p_c$ fields. 
 
Once we apply the velocity correction in region $\mathcal{T}_i$, the divergence-free condition for the velocity is violated 
in the rest of the simulation domain and equation~\ref{eq:continuity} is valid only if we define 
$\vb{U} = \vb{\hat{U}} - \grad{\Phi}$, where $\Phi$ is a 
scalar field. 
From equation~\ref{eq:continuity}, $\laplacian{\Phi} = \grad \vdot \vb{\hat{U}}$, which is equivalent to 
apply a force-term
to equation~\ref{eq:inc-Navier}. This approach and its potential to resolve the motion of particles in liquids 
has been extensively discussed by Kloss \emph{et al.} \cite{kloss_2012}.

\subsection{DEM Method}\label{subsec-DEM}

The Discrete Element Method was first presented in 1979 by Cundall \cite{cundall_1979}. One of the main 
characteristics of DEM simulations is its efficacy when resolving the granular medium at the particle scale. 
The DEM is a Lagrangian method, meaning that all particles in the computational domain have their trajectories 
solved explicitly in every time step. Thus, the DEM is capable of simulating a wide range of phenomena, 
such as dense and dilute particulate systems, as well as
rapid and slow granular flow. Here, we briefly discuss the main features of the model, and a more detailed 
description of the method was published by Goniva \emph{et al.} \cite{goniva_2012}.

For a particle $i$ sedimenting under gravity in a liquid, 
its trajectory is calculated using the following force and torque balances:

\begin{equation}\label{eq:DEM_forces}
 m_i \vb{\ddot{x}}_i = m_i \vb{g} + \vb{F}_{i, f} + \sum_{j = 1}^{N_{p}} \vb{F}_{i, j} + \sum_{N_{w}} \vb{F}_{i, w}
\end{equation}

\noindent and

\begin{equation}\label{eq:DEM_torques}
 I_i \frac{d \vb*{\omega}}{dt} = \sum_{j = 1}^{N_{p}} \vb{r}_{i, j}  \cross \vb{F}_{i, j}^{\perp} + \vb{T}_{i, f}
\end{equation}

where $i$ represents the index of the particle, $m_i$ is its mass, $\vb{x}_i$ its position vector and $I_i$ its 
moment of inertia. $N_p$ represents the number of neighboring particles, and $N_w$ the number of near walls. 

In Table~\ref{table-DEM_forces} we describe the forces and torques from 
equations~\ref{eq:DEM_forces} and~\ref{eq:DEM_torques}. The particle-particle $\vb{F}_{i, j}$ and the 
particle-wall $\vb{F}_{i, w}$ contact forces are written in terms of their ortoghonal and parallel components with
respect to the plane of contact.

In the method originally presented in \cite{hager_2012},
the buoyant force was explicitly accounted for in the total fluid force acting on the particles.
The current implementation explicitly considers the gravitational field acting on the fluid phase. 
Thus, we represent the fluid-particle interaction force by the last equation in 
Table~\ref{table-DEM_forces}. It generalizes the original approach 
presented by Shirgaonkar \emph{et al.} \cite{shirgaonkar_2009}.
Importantly, the new approach includes a multiplicative term ($1 - \phi_c$) in the pressure gradient accounting for 
the shape of the particle. 
Note, the void fraction field $\phi_c$ represents the fluid occupation in the volume element $c$. 

\begin{table}
\centering
\caption{Summary of forces and torques acting on the particle. For a full description of the contact forces, 
refer to \cite{goniva_2012}.}
\label{table-DEM_forces}       % Give a unique label
% For LaTeX tables you can use
\begin{tabular}{l l}
\hline
Description & Interactions \\ \hline 
Particle-particle contact force $\vb{F}_{i, j}$ & $\vb{F}_{i, j}^{\bot} + \vb{F}_{i, j}^{\parallel}$ \\

Particle-wall contact force $\vb{F}_{i, w}$ & $\vb{F}_{i, w}^{\bot} + \vb{F}_{i, w}^{\parallel}$ \\ 

\\

Particle-fluid torque $\vb{T}_{i, f}$ & 
$\begin{aligned}
    \sum_{c \in \mathcal{T}_i} & (\vb{x}_c - \vb{x}_i) \cross \vb{F}_{i, D}
\end{aligned}$ \\ 

\\
 
Fluid-particle force $\vb{F}_{i, f}$ \cite{shirgaonkar_2009} & 
$\begin{aligned}
    \sum_{c \in \mathcal{T}_i} & V_c [\mu \laplacian \vb{U}_c - \\ 
    & \grad{p} \, (1 - \phi_c)] 
\end{aligned}$ \\ 
 \hline

\end{tabular}
% Or use
%\vspace*{5cm}  % with the correct table height
\end{table}

\subsection{CFD-DEM coupling}\label{subsec-CFD-DEM}

The coupling procedure can then be described by the following steps:
\begin{enumerate}[-]
 \item Determine the position and velocity of each particle in the domain from the DEM state.
 \item Assign the particles to the cells that contain their centers. 
 \item Solve the incompressible Navier-Stokes equations for the fluid phase using the PISO algorithm.
 \item Correct the velocity found in the previous step in the grid cells where particles are located.
 \item Calculate $\vb{F}_{i, D}$ for all particles, this force is applied in the next set of DEM time steps. 
 \item Lastly, the boundary conditions are applied in the fluid phase, and we return to the first step.
\end{enumerate}

\subsection{Simulation Details}\label{subsec-simulation_details}
Here, we simulate the motion of a single particle in a 3D viscous fluid. 
The radius and density of the particle are $R = d/2 = 1 \, mm$ and 
$\rho_p = 7850 \, kg / m^3$, respectively. 
The fluid has density $\rho_f = 970 \, kg/m^3$ and several kinematic 
viscosities are explored $\nu = [1000, 2000, 5000, 12500]$ in cSt. 
Specifically, we examine a sphere left to settle from rest in a quiescent fluid.
In this scenario, we expect the particle to settle with terminal velocity 
$v_{S} = \frac{2}{9 \mu} (\rho_p - \rho_f) g R^2$ and characteristic time $\tau = \frac{2}{9 \mu} \rho_p R^2$.
We set a grid resolution equal 
to $\Delta x = d/8$ and CFD and DEM time steps $\Delta t = \tau / 35$.
In all cases, the total simulation time was $20 \tau$ seconds and the Reynolds numbers were inferior to $0.1$.

\subsection{Sphere settling between plane parallel walls}\label{subsec-particle_near_wall}

Complementarily, we study the settling terminal velocity of a sphere under confined conditions.
Back in 1923, Fax\'en studied the effect of plane parallel walls on reducing the settling velocity of spheres 
\cite{faxen_1923}. In a low Reynolds number regime, for a sphere centered 
between two plane parallel walls 
separated by a distance $H$, Fax\'en proposed that the drag force applied to the particle by:
\begin{equation}\label{eq:faxen}
 F_z = \frac{6 \pi \mu R v}{1 - 1.004 s + 0.1475 s^2 - 
 0.131 s^4 - 0.0644 s^5} 
\end{equation}
where $s = \frac{2R}{H} = \frac{d}{H}$, $R$ the sphere radius and $v$ the settling velocity. 
This expression approximates the actual force acting on the sphere when it is located far from the walls ($s^{-1} > 2$). 
A more general solution for the drag force was later introduced by Ganatos \emph{et al.}, covering 
configurations that were out of the validity of Fax\'en's solution \cite{ganatos_1980}.

\section{Results and discussions}\label{sec-results}

\subsection{Validation}\label{subsec-validation}

As a first step, we explore a system with dimensions $10d \times 10d \times 25d$. 
Initially, the particle is centered in the xy-plane at the location of
$z = 17.5d$. Then, it is left to settle from rest, it accelerates and reaches its terminal velocity. 
%\textbf{Note that comparing to Eq.~\ref{eq:faxen}, we assume the impact of confinement on the drag coefficient is negligible, given the large system size.}

\begin{figure}
\centering
\includegraphics[width=8cm,clip]{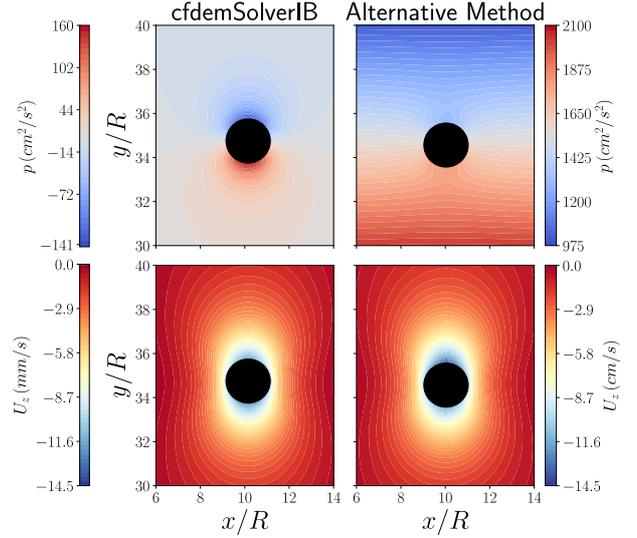}
\caption{Comparison of $p / \rho_f$ and $U_z$ fields for our approach and cfdemSolverIB \cite{hager_2014}. 
a) and b) represent the pressure field, c) and d) the velocity field $U_z$. }
\label{fig-fields}       % Give a unique label
\end{figure}

Figure~\ref{fig-fields} illustrates the system state obtained after $20\tau$ seconds, using both approaches.
It compares the kinematic pressure (Figs.~\ref{fig-fields}a and ~\ref{fig-fields}b) and velocity 
(Figs.~\ref{fig-fields}c and ~\ref{fig-fields}d) fields. 
Note that the cfdemSolverIB approach (Fig.~\ref{fig-fields}a) does not account for the pressure gradient imposed by the external
field in the fluid. It is also noticeable that the kinematic pressure perturbation of the particle movement is 
significantly smaller than the hydrostatic pressure gradient (Fig.~\ref{fig-fields}b). 
Remark that when using cfdemSolverIB, the local pressure field is solely defined 
by the corresponding variations in the velocity field,  as a result of the momentum balance equations. 
In our approach, apart from that local pressure contribution, we also have the gravitational field acting on the fluid.
However, no relevant impact is detected in the velocity fields (Figs.~\ref{fig-fields}c and ~\ref{fig-fields}d).

\begin{figure}
\centering
\includegraphics[width=8cm,clip]{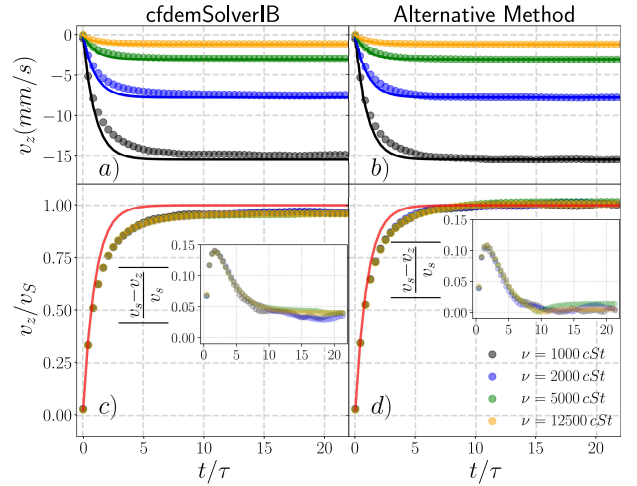}
\caption{a) and b) Time evolution of the settling velocity of the sphere for both methods. 
c) and d) Time evolution of the settling velocity scaled by the corresponding Stokesian terminal velocity. 
The insets represent the relative differences in each case.
Note that a more accurate asymptotic limit velocity is reached in our method. 
All solid lines represent the expected Stokesian dynamics.
}
\label{fig-viscosity-comparison}       % Give a unique label
\end{figure}

Next, we investigate the stability of our method when computing %different
very low Reynolds number conditions.
Thus, we perform a systematic study, running a set of distinct scenarios increasing the fluid viscosity $\nu$. 
Fig.~\ref{fig-viscosity-comparison}a and Fig.~\ref{fig-viscosity-comparison}b illustrate the outcomes obtained
using both approaches, in comparison with the analytical prediction of Stokes' law. 
Our results indicate that both methods yield a reasonable description of the settling process and the particle
reaches a viscous-dependent terminal velocity $v_S^{'}$ at the long-time limit.
As expected, the results obtained using both methods collapse in a single curve using 
$v_{S} = \frac{2}{9 \mu} (\rho_p - \rho_f) g R^2$ and $\tau = \frac{2}{9 \mu} \rho_p R^2$. 
It is noticeable that both approaches slightly fails to reproduce the analytical solution during the initial 
acceleration process (see the insets in Figs. ~\ref{fig-viscosity-comparison}c and ~\ref{fig-viscosity-comparison}d).

However, assuming a negligible impact of the finite size boundary conditions implemented 
(in line with the results introduced in \cite{ganatos_1980}), the present approach compares better with the asymptotic 
limit behavior. We speculate that the difference arises from the impact of dynamic pressure gradients acting on the 
particle surface for the original particle discretization. Hence, when a volumetric correction is applied to compute the 
pressure gradients to calculate the force $\vb{F}_{i, f}$, we account for the \emph{local} solid fraction, resolving the 
corresponding pressure gradient more accurately.

%However, we obtain that the present approach compares better with the asymptotic limit behavior, assuming the impact of confinement on the drag coefficient is negligible, given the large system size.We could argue that the differences arise from the following factor. In the previous method, the impact of the dynamic pressure gradients on the surface of the particle is overestimated, given the finite discretized system. In contrast, following the new approach a volumetric correction is applied when computing the dynamic buoyancy from the pressure gradients.Thus, computing the force $\vb{F}_{i, f}$, we account for the actual solid fraction of the site, applying the corresponding pressure gradient accurately.

\subsection{Neighboring walls effect}\label{subsec-results}

Complementarily, we examine the impact of the confinement on the dynamics of a settling particle.
Specifically, we explore the behavior of a particle located between two parallel walls. 
The particle is centered in the xy-plane of a box with dimensions $10d \times Hd \times 25d$ and is left to settle from rest.
We systematically vary the distance between the walls $H/d = [1.2, 1.4, 1.8, 2.6]$, in terms of particle diameter. 

\begin{figure}
\centering
\includegraphics[width=8cm,clip]{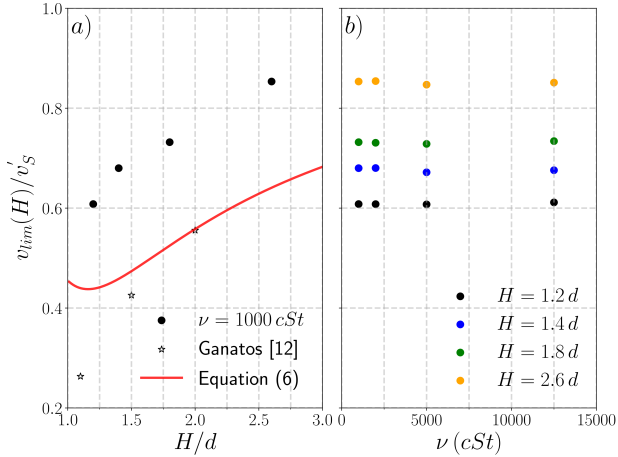}
\caption{a) Settling velocity of a sphere between two planes scaled by the terminal velocity in a large system when
varying the distance between walls $H$ for a liquid with $\nu = 1000 \, cSt$. 
The solid red line represents the inverse of the denominator of Equation \ref{eq:faxen}. 
b) When analyzing the same velocities ratios of a) varying the viscosity, we observe a similar pattern for the
confinement effect on the terminal velocity.}
\label{fig-near-wall}       % Give a unique label
\end{figure}

Fig~\ref{fig-near-wall}a shows the terminal velocity of the sphere moving between two walls and obtained for 
different confinement conditions, fixing the fluid viscosity $\nu = 1000 \, cSt$. 
For simplicity sake, the data values are rescaled using the terminal velocity $v_S^{'}$ that corresponds 
to the infinite system size limit. 
We observe that the simulations can reproduce a decrease in the terminal velocity when reducing the distance 
between the confining walls.
Although, the sphere settles with a velocity larger than the analytical solutions provided by
Fax\'en \cite{faxen_1923} and Ganatos \emph{et al.} \cite{ganatos_1980}.
Meanwhile, Fig~\ref{fig-near-wall}b shows that our implementation has a consistent behavior for the 
range of viscosities investigated here. A similar result was previously obtained for a sphere in a creeping flow 
condition using an incompressible method based on OpenFOAM software \cite{ponianev_2016}.  

\section{Conclusions}\label{sec-conclusions}

We introduce an alternative methodology to solve the particle-fluid interaction in the 
\textit{resolved} CFDEM{\textregistered}coupling framework. 
The new approach explicitly accounts for the body force acting on the fluid phase, 
when computing the local momentum balance equations. 
The proper representation of the pressure profile in the 
fluid phase led to a single force model that no longer requires an explicit buoyant force 
to account for the gravitational interaction. 
Our method was able to outperform the cfdemSolverIB method for the settling velocity of a sphere, 
exhibiting more accurate results. 

Complementarily, we study the dependence of the settling velocity of a sphere under confined conditions.
The results indicated a consistent velocity decrease when reducing the distance between confining walls. 
Nevertheless, our numerical outcomes did not reproduce the analytical solutions found in the literature.   \\ 

\acknowledgement{
This work has been partially supported by Fundaci\'on Universidad de Navarra and the project from FIS 2017-84631 (MINECO). 
I. Fonceca also thanks Asociaci\'on de Amigos de la Universidad de Navarra for his scholarship.
}

\end{document}